\documentclass[manuscript]{aastex}

\slugcomment{Not to appear in Nonlearned J., 45.}

\shorttitle{Analysis on Mechanisms of Reconnection 
            Rate Enhancement}
\shortauthors{Wang et al.}

\begin{document}

\title{Three-dimensional MHD Magnetic Reconnection 
       Simulations with Finite Guide Field: Proposal 
       of the Shock-Evoking Positive-Feedback Model}

\author{Shuoyang Wang and Takaaki Yokoyama}
\affil{Department of Earth and Planetary Science, 
       The University of Tokyo, 7-3-1 Hongo, 
       Bunkyo-ku, Tokyo, 113-0033, Japan }
\email{wangshuoyang@eps.s.u-tokyo.ac.jp}

\and

\author{Hiroaki Isobe}
\affil{Shishukan, Kyoto University,
       Yoshidanakaadachicho, Kyoto Sakyo-ku, 
       Kyoto 606-8306, Japan}

\begin{abstract}

Using a three-dimensional magnetohydrodynamic 
model, we simulate the magnetic reconnection 
in a single current sheet. We assume a finite 
guide field, a random perturbation on the 
velocity field and uniform resistivity.

Our model enhances the reconnection rate 
relative to the classical Sweet-Parker model 
in the same configuration. The efficiency of 
magnetic energy conversion is increased by 
interactions between the multiple tearing 
layers coexisting in the global current sheet. 
This interaction, which forms a 
positive-feedback system, arises from coupling 
of the inflow and outflow regions in different 
layers across the current sheet. The coupling 
accelerates the elementary reconnection events, 
thereby enhancing the global reconnection rate. 
The reconnection establishes flux tubes along 
each tearing layer. Slow-mode shocks gradually 
form along the outer boundaries of these tubes, 
further accelerating the magnetic energy 
conversion. Such positive-feedback system is 
absent in two-dimensional simulation, 
three-dimensional reconnection without a guide 
field and a reconnection under a single 
perturbation mode. We refer to our model as 
the ``shock-evoking positive-feedback" model.
\end{abstract}


\section{Introduction}

Magnetic reconnection is considered to source 
the rapid conversion of magnetic energy in 
various solar coronal activities with 
extremely high Lundquist number 
$S = L v_{A}/\tilde{\eta} \sim 10^{14}$, 
where $L$, $v_{A}$ and $\tilde{\eta}$ denote 
the current sheet length, Alfv\'{e}n speed and 
magnetic diffusivity, respectively. The two 
classical reconnection models are the 
Sweet-Parker model \citep{swe58,par63} 
and Petschek's model \citep{pet64}. 
In the Sweet-Parker model, the reconnection 
rate scales as 
$v_{inflow}/v_{A} \sim 1/\sqrt{S}$, several 
orders of magnitude slower than required in 
many solar and astronomical applications. 
The Petschek's model yields a sufficiently 
fast reconnection rate ($\sim 1/\ln{S}$) 
by virtue of the localized diffusion region 
and the extended slow-mode shocks. The localized 
diffusion region is thought to originate from 
microscopic plasma processes occurring on scales 
many orders of magnitude smaller than the global 
scale. Thus, it is necessary to address the huge 
scale gap between the micro and the global 
scales while maintaining efficient global energy 
conversion.

\citet{bis86} argued that Sweet-Parker sheets with 
aspect ratio (i.e., length to thickness) exceeding 
$100$ are vulnerable to secondary tearing instability. 
\citet{shi01} developed a fractal reconnection model 
based on this concept. They reasoned that once 
plasmoids are formed by the primary tearing 
instability, plasmoid ejection and growth stretch the 
intervening current sheets, increasing the aspect 
ratio of the current sheets. Eventually these 
current sheets become unstable to secondary tearing, 
and disintegrate into chains of smaller-scale 
plasmoids and current sheets. By this stepwise process, 
an initially long and thick current sheet can reduce 
to the scale of the ion Larmor radius or ion inertial 
length. This hierarchical structure can plausibly 
couple the largest and smallest scales. High-resolution 
numerical simulations have proved the feasibility of 
this scheme \citep{bar11} and also showed that the 
reconnection rate becomes independent of the Lundquist 
number when $S \gtrsim S_{c} \sim 10^4$ 
\citep[e.g.,][]{lou12}.

This fractal reconnection model assumes translational 
invariance in the direction perpendicular to the 
reconnection plane. This assumption is violated in 
the presence of three-dimensional (3D) instabilities, 
such as the kink-like instability reported by 
\citet{dah92}. These authors found fast growth and 
saturation of the tearing mode along the anti-parallel 
magnetic field. Meanwhile, the oblique modes (with 
a component perpendicular to the tearing plane) 
continue growing and dominate in the later phase. 
Moreover, in the absence of a guide field, this 
kink-like instability will suppress the coalescence 
instability in the tearing layer, completely breaking 
the initial laminar structure \citep{dah02}. This 
instability is expected to affect the fractal 
reconnection in 3D scenario.

Another concern is the emergence of multiple 
tearing layers \citep{gal77}. When a sheared current 
sheet is subjected to a multi-modal perturbation 
(comprising modes ${\bf k_1, k_2, ..., k_n}$), 
multiple layers whose local magnetic field 
orientation satisfy ${\bf k} \cdot {\bf B} = 0$ can 
emerge and are expected to interact when they become 
sufficiently close. Interaction should disrupt the 
laminar structures of the layers, altering their 
internal fractal reconnections. \citet{ono04} studied 
the nonlinear evolution of several tearing layers 
coexisting inside a sheared current sheet in an 
incompressible plasma. They found that although the 
two-dimensional (2D) mode grows more rapidly than the 
emerging modes in a linear analysis, the emerging 
modes are oblique modes at an early stage. The 
resulting energy cascade gradually extends outwards 
from the current sheet center, leading to a final 
turbulent state. Moreover, the inverse energy 
transfer implies coalescence of the magnetic islands. 
\citet{lan08} further confirmed the three-dimensionality 
of the initially emerging modes in a compressible 
plasma simulation. Thus, the preferred source of plasmoid
instability growth is changed and the reconnection 
enhancement concept should be modified in the 3D case. 
But neither \citet{ono04} nor \citet{lan08} explicitly 
discussed the changes in the reconnection rate. For this 
purpose, we seek more details of the magnetic energy 
consumption.

This paper presents a detailed study of magnetic 
reconnection under tearing instability in a 3D resistive 
compressive MHD environment with a finite guide field. 
In Section 2, we introduce the simulation model. 
In Section 3 and 4, simulation result and its discussion 
are presented. Section 5 makes a summary of the whole 
study.

\section{Simulation Model}

The 3D resistive MHD equations are solved in 
Cartesian coordinates. Viscosity, gravity and heat 
conduction are neglected for simplicity. The basic 
equations are as follows:

\begin{equation}
\frac{\partial \rho}{\partial t}
  +({\bf v} \cdot \nabla) \rho
  =-\rho (\nabla \cdot {\bf v})
\end{equation}
\begin{equation}
\rho \left( \frac{\partial {\bf v}}{\partial t}
  +({\bf v} \cdot \nabla ){\bf v} \right)
  =-\nabla p+\frac{{\bf J}\times {\bf B}}{c}
\end{equation}
\begin{equation}
{\bf J}=\frac{c}{4\pi }\nabla \times {\bf B}
\end{equation}
\begin{equation}
\frac{\partial p}{\partial t}
  +({\bf v}\cdot \nabla )p
  =-\gamma p(\nabla \cdot {\bf v})
   +(\gamma -1)\eta {\bf J}^{2}
\end{equation}
\begin{equation}
\frac{\partial {\bf B}}{\partial t}
  =\nabla \times \left( {\bf v} \times 
   {\bf B} - c \eta {\bf J} \right)
\end{equation}
\begin{equation}
p=\frac{\rho}{\bar {m}} k_{B}T.
\end{equation}
Here, $k_{B}$ is the Boltzmann constant, 
$\eta$ is the resistivity, $\gamma = 5/3$, 
$c$ is the speed of light, and $\bar{m}$ is 
the mean particle mass. For a fully ionized 
hydrogen gas, $\bar{m} = 0.5m_{p}$, where 
$m_{p}$ is the proton mass. The other 
variables take their usual meanings.

All quantities are normalized by their 
characteristic values. The length scale is 
the width of the initial current sheet 
($=\delta$). The time scale is normalized by 
$t_{A} = \delta /v_{A0}$, where $v_{A0}$ 
is the Alfv\'{e}n velocity. The initial mass 
density $\rho_{0}$ normalizes the mass 
density. The magnetic field is normalized 
by $B_{0}=v_{A0}\sqrt{\rho_{0}}$. Consequently, 
the current density is normalized by 
$J_{0} = cB_{0}/\delta$ and the plasma 
pressure is normalized by $p_{0} = B_{0}^{2}$.

\begin{figure}[!h]
    \centering
    \includegraphics[scale=1.]{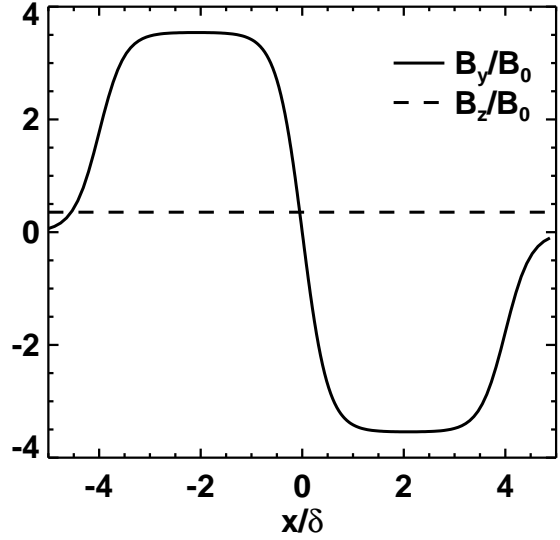}
    \caption{Initial magnetic field plot along 
             $y = 0, z = 0$. Solid line is the 
             normalized $B_y(x,y=0,z=0)$ and 
             dashed line is the normalized 
             $B_z(x,y=0,z=0)$.
  \label{fig01}}
\end{figure}

The initial magnetic field comprises an 
anti-parallel component $B_{y}(x)$ and a uniform 
finite guide field $B_{z}$, as plotted in 
Fig.$~$\ref{fig01}:
\begin{eqnarray}
{\bf B}& = & B_{y}{\bf e_{y}}+B_{z}{\bf e_{z}} \nonumber \\
       & = & B_{y0}\tanh \left( \frac{x}{0.5\delta} \right) 
             \left\{ \frac{1}{2} \bigg[ \tanh \left( 
             \frac{\mid x \mid -4\delta}{0.5\delta} \right) 
             -1 \bigg] \right\} {\bf e_{y}} + \alpha B_{y0} 
             {\bf e_{z}}
\end{eqnarray}
where $B_{y0} = \sqrt{4\pi}B_{0}$ and 
$\alpha$ controls the magnitude of the guide 
field $B_{z}$. Here we adopt $\alpha = 0.1$ 
in the basic model. A magnetic field with 
non-zero $\alpha$ shears the structure along 
the $x$-axis. 
The magnetic diffusivity is assumed to be 
spatially and temporally constant with a 
magnitude of 
$\tilde{\eta} = c^{2} \eta / (4 \pi) \sim 
3 \times 10^{-4} \delta^{2} / t_{A}$. 
Resulting Lundquist number is thus 
defined by the Alfv\'{e}n speed 
$v_{A} = v_{A0}\sqrt{1+\alpha^{2}}$, 
$S = v_{A}L_{y}/(2\tilde{\eta}) \sim 2.6 \times 10^{5}$.

The simulation begins with a uniform mass 
density $\rho_{0}$ throughout the simulation 
space. Pressure balance between the gas and 
magnetic field is also maintained in the 
whole box, giving:
\begin{equation}
p+\frac{B^2}{8\pi} 
  = \frac{B_{0}^2}{2}(1+\alpha^2)(1+\beta) 
\end{equation}
in which $\beta$ is the ratio between the 
plasma and magnetic field pressures at the 
minimum plasma pressure; we set $\beta = 0.2$.

To initiate the reconnection, we add a 
random velocity perturbation with small 
amplitude 
($v_{x}, v_{y}, v_{z} \lesssim 1\times 
10^{-3} v_{A0}$) over the whole 
simulation domain. Short-wavelength 
perturbations are eliminated. The 
remaining waves satisfy 
$1/k_y^2+1/k_z^2 \geqslant 116 \delta^2$, 
also limited by 
$\mid k_y \delta \mid \leqslant 2$ and 
$\mid k_z \delta \mid \leqslant 2$.

The simulation box size 
($L_{x} \times L_{y} \times L_{z}$) is 
$10\delta \times 24\delta \times 6\delta$, 
containing $240\times768\times192$ grids. 
To resolve the tearing layer, we construct 
non-uniform grids with 
$\Delta x \geqslant 0.02 \delta$ along 
the $x$-direction. Uniform grids with 
$\Delta y = \Delta z = 0.03125 \delta$ are 
constructed in the $y$- and $z$-directions. 
Periodic boundary conditions are imposed on 
each border. The calculations are performed 
in CIP-MOCCT code \citep{kud99} with an 
artificial Lapidus viscosity \citep{lap67}.

\section{Simulation results}

The efficiency of the magnetic energy 
conversion is approximately five times higher 
in our 3D reconnection than in the Sweet-Parker 
reconnection. We propose a detailed mechanism 
of reconnection enhancement, referring to our 
new model as the 
``shock-evoking positive-feedback" model.

\subsection{Global structure}

\begin{figure}[!h]
    \centering
    \includegraphics[scale=1.0]{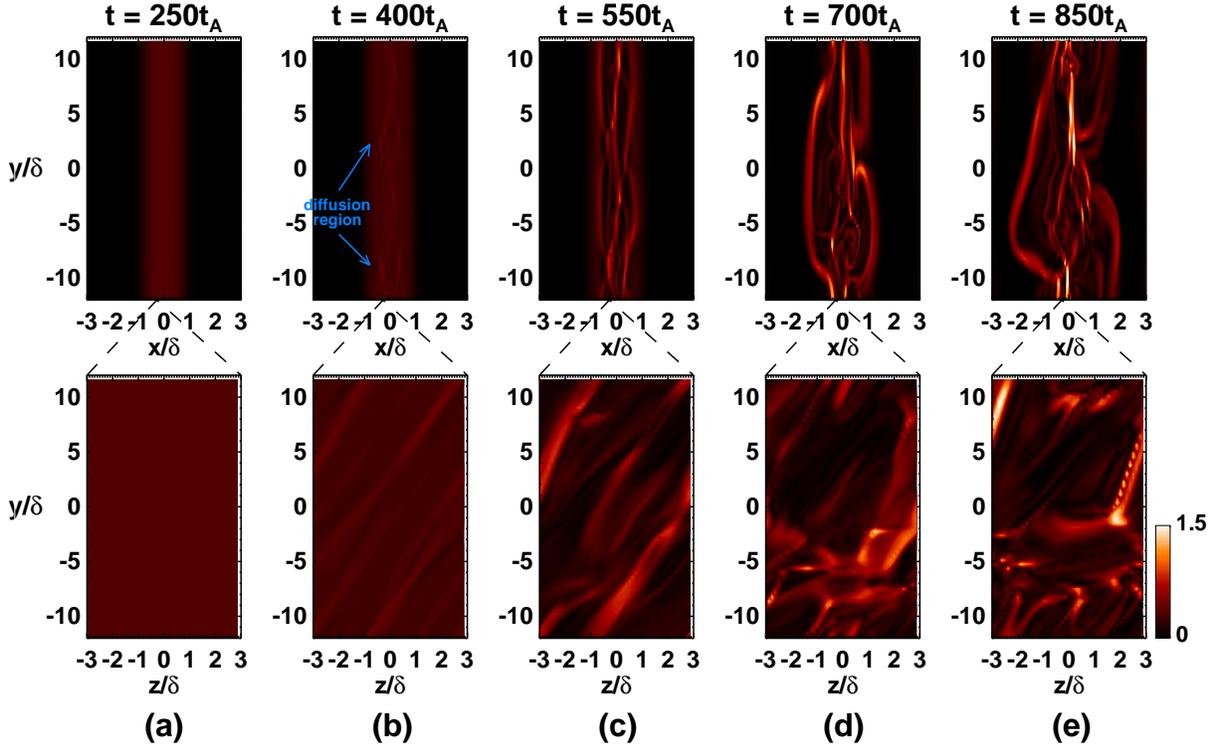}
    \caption{Current density $|{\bf J}|/J_{0}$ 
             in the $z = 0$ plane (upper panel) 
             and in the $x = -0.3\delta$ plane 
             (lower panel).}
    \label{fig02}
\end{figure}

\begin{figure}[!h]
    \centering
    \includegraphics[scale=1.0]{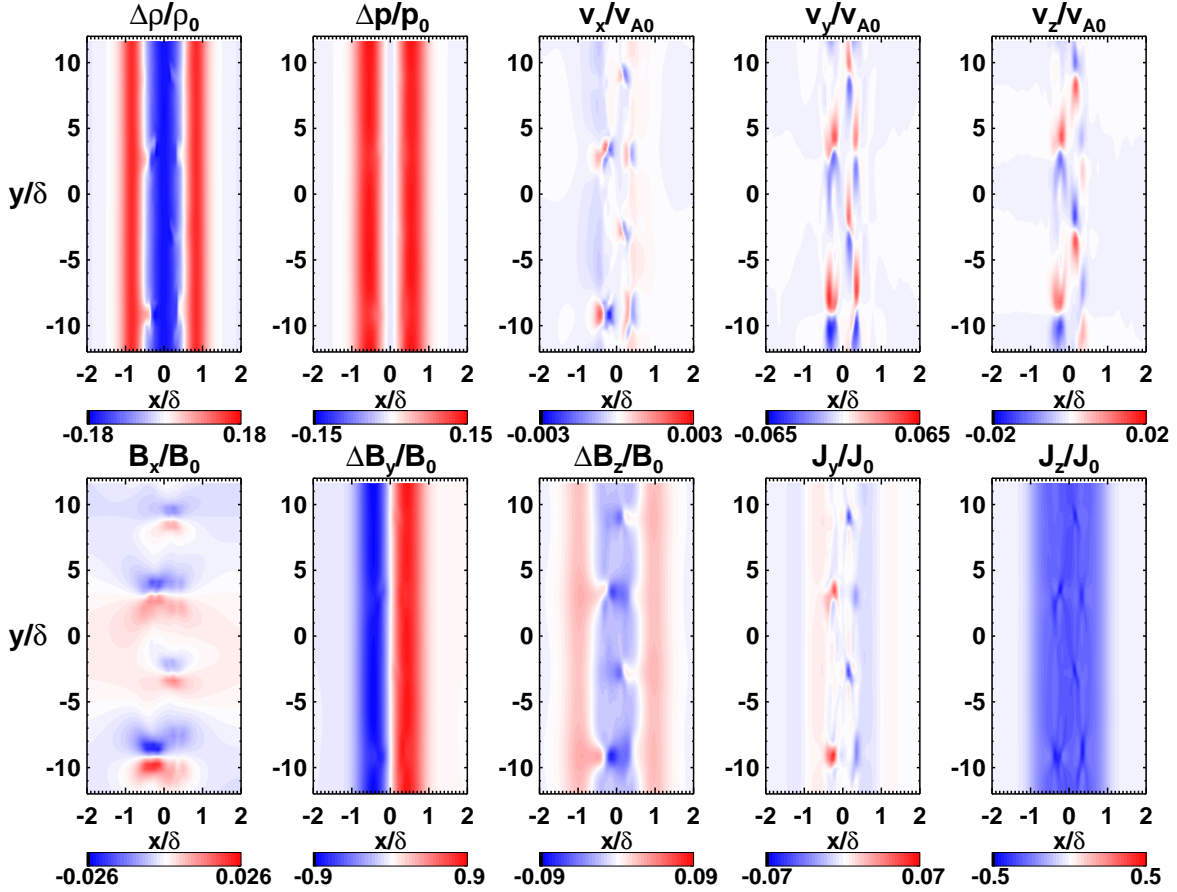}
    \caption{Global patterns of variables on 
             the $z = 0$ plane at $t = 400t_{A}$; 
             plasma density difference ($\Delta \rho$), 
             pressure difference ($\Delta p$), 
             velocity components ($v_x, v_y, v_z$), 
             magnetic field components ($B_x$) 
             and differences 
             ($\Delta B_{y}, \Delta B_{z}$), 
             current density component ($J_y, J_z$)}
    \label{fig03}
\end{figure}

\begin{figure}[!h]
    \centering
    \includegraphics[scale=1.0]{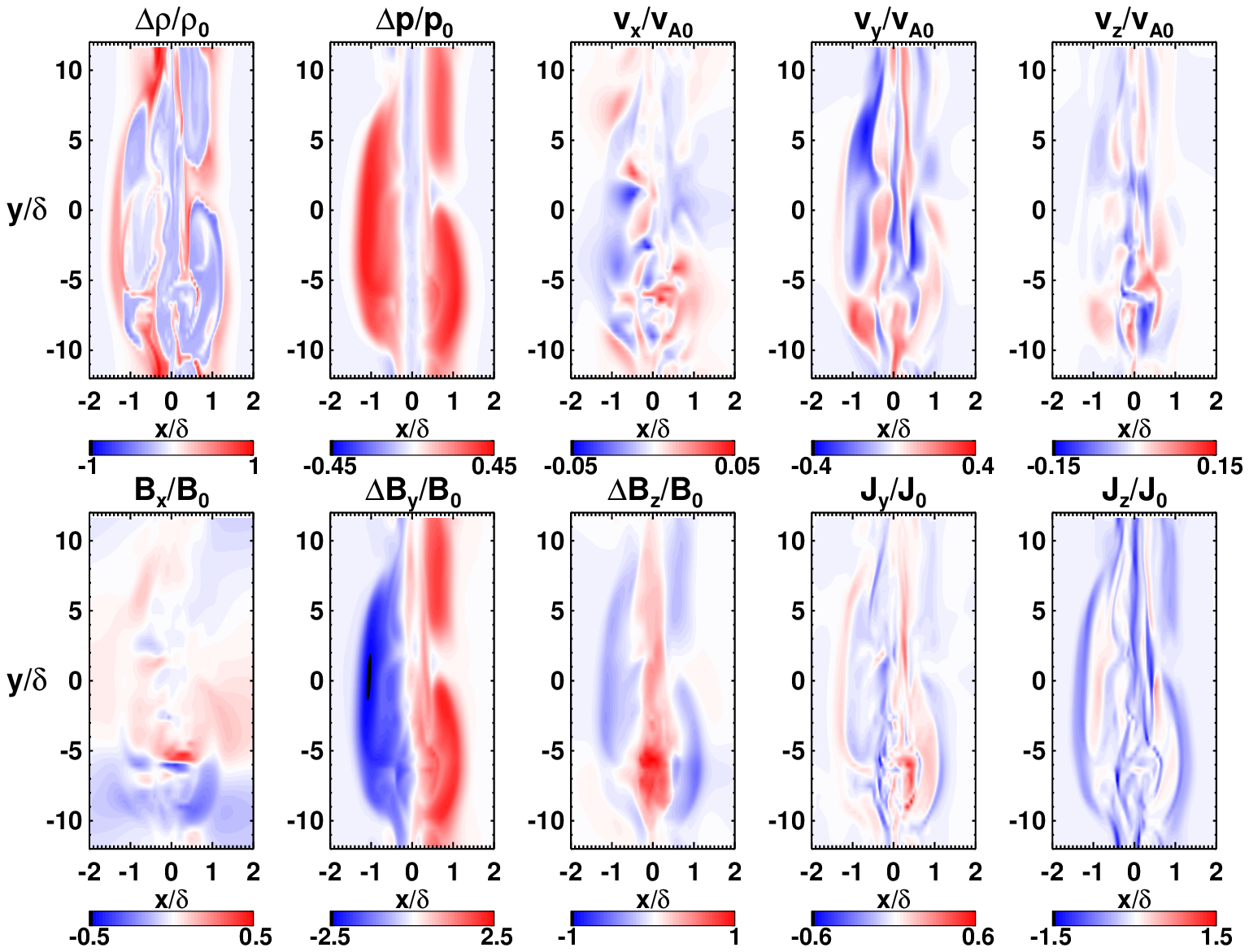}
    \caption{Global patterns on the $z = 0$ plane 
             at $t = 700t_{A}$. Variables are 
             defined in the same way as 
             Fig.$~$\ref{fig03}.}
    \label{fig04}
\end{figure}

The upper and lower panels of Fig.$~$\ref{fig02} 
show the temporal evolution of the current 
density $|{\bf J}|/J_{0}$ in the $xy$-plane 
($z = 0$) and the $zy$-plane ($x = -0.3\delta$), 
respectively. In the early phase 
(Fig.$~$\ref{fig02}(a)), diffusion dominates 
and no clear pattern is observed inside the 
primary (central) current sheet. Gradually, 
tearing instability \citep{fur63} emerges and 
two chains of smaller diffusion regions (or 
reconnection sites) develop and grow adjacent 
to $x = \pm 0.3 \delta$ (Fig.$~$\ref{fig02}(b)). 
Within these chains, the current density is 
much higher than the surroundings. The chains 
form a web-like pattern across the global 
current sheet. Fig.$~$\ref{fig03} and 
Fig.$~$\ref{fig04} present the plasma density, 
pressure, velocity, magnetic field, and current 
density at $t = 400t_{A}$ and $700t_{A}$ respectively, 
in the $z = 0$ plane. The $\Delta$ notation 
defines a difference; for example, $\Delta \rho$ 
is defined as $\Delta \rho = \rho-\rho_{initial}$, 
where $\rho_{initial}$ is the initial value of 
$\rho$. The structures are well-organized in 
each plot of Fig.$~$\ref{fig03}, but becomes 
distorted by flows in the later phase 
(Fig.$~$\ref{fig04}).

Oblique lines (diffusion lines) are formed by 
the diffusion regions in the $zy$-plane, as 
shown in the lower panel of Fig.$~$\ref{fig02}(b). 
The positions of these diffusion lines along 
the $x$-axis are determined by resonance layers 
(resonant tearing layers), which arise from 
the periodic boundary conditions in the $y$- 
and $z$- directions. This slab structure 
resembles the cylindrically symmetric rational 
surface in a tokamak \citep[e.g.,][]{bel06}. 
The safety factor $q$, which denotes the 
period of the magnetic field lines cycling 
(sheet-wise) across the $zy$-plane, is calculated 
along the $x$-axis as follows:
\begin{equation}
q\left( x \right) = \frac{L_{y}}{L_{z}} 
      \left| \frac{\langle B_{z} \rangle_{x}}
      {\langle B_{y} \rangle_{x}} \right|.
\end{equation}
Here, $L_{y}$ and $L_{z}$ are the box sizes 
in the $y$- and $z$-directions, respectively. 
$\langle B_{y} \rangle_{x}$ and 
$\langle B_{z} \rangle_{x}$ are the mean 
magnetic field components on the $zy$-plane 
at the corresponding point $x$. Strictly 
speaking, provided that the $q$ value in a 
$zy$-plane satisfies 
\begin{equation}
q = \frac{m}{n}
\end{equation}
where $m$ and $n$ are integers, the wave signal 
is strengthened and the layer develops into 
a resonance layer. Clearly numerous layers are 
expected. However, due to the different growth 
rates of these modes, a limited number of layers 
can resonate locally. When $m$ and $n$ are large, 
the corresponding wavelength is small; 
hence, the instability is easily suppressed by 
magnetic tension force.

\begin{figure}[!h]
    \centering
    \includegraphics[scale=1.05]{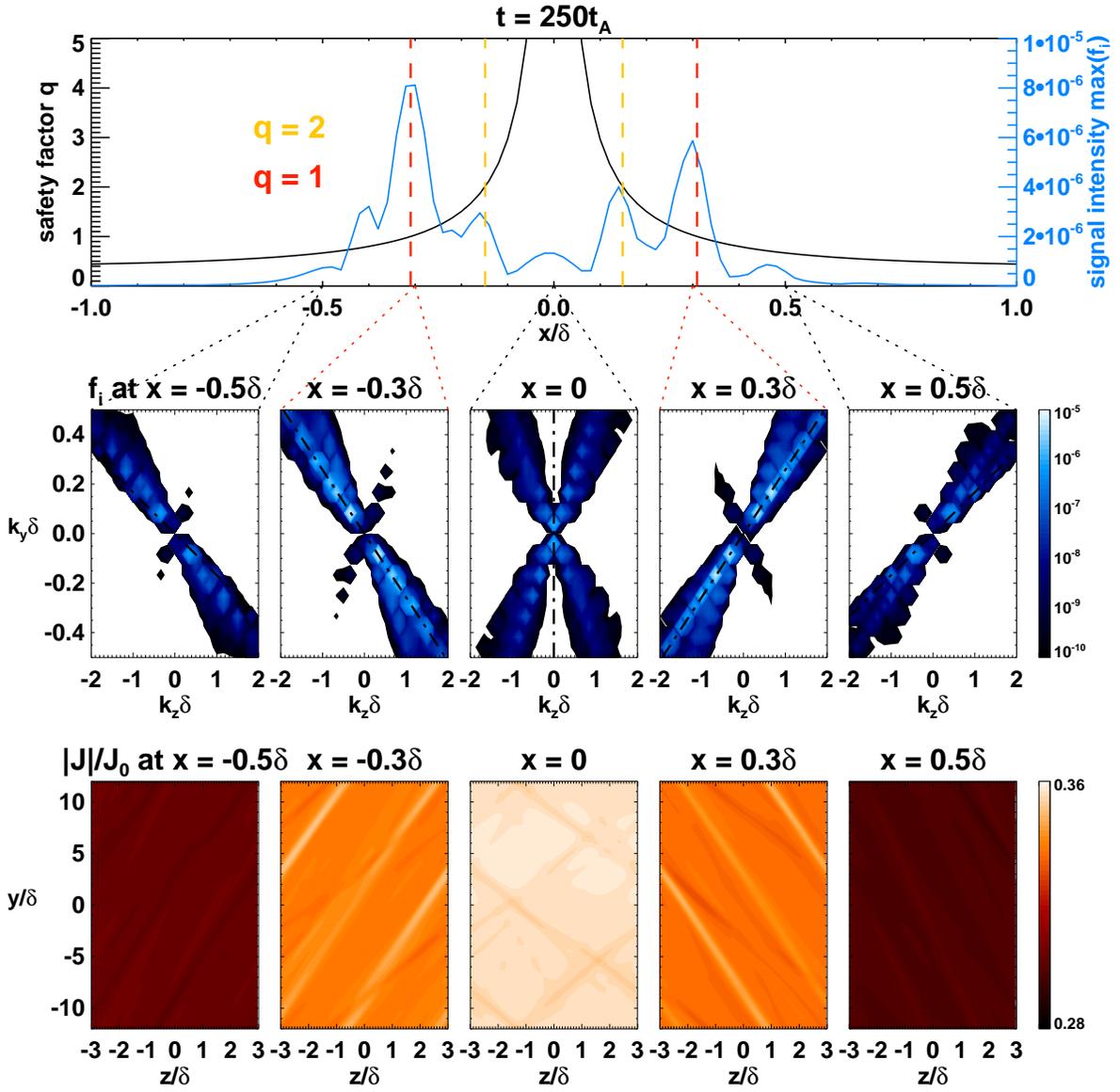}
    \caption{Upper panel: safety factor $q$ and 
             local maximum $\max_{k_z,k_y}(f_i)$ 
             at points along the $x$-axis. 
             Middle panel: $f_i$ in $k$-space in 
             selected $zy$-planes.
             Lower panel: normalized $|{\bf J}|$ 
             in the corresponding layers.}
    \label{fig05}
\end{figure}

The safety factors calculated by Eq.$~$(9) 
at $t = 250t_{A}$ are plotted as black solid 
line in the upper panel of Fig.$~$\ref{fig05}. 
To find the most unstable mode in the box, the 
Fourier transformation of the current density 
$|{\bf J}|/J_{0}$ on each $zy$-plane is 
calculated along the $x$-axis. The signal 
intensity is calculated as 
\begin{equation}
f_{i}(x,k_{z},k_{y}) 
     = \left| \int_{-\frac{L_{z}}{2}}^{\frac{L_{z}}{2}}
              \int_{-\frac{L_{y}}{2}}^{\frac{L_{y}}{2}}
               [|{\bf J}(x,y,z)|/J_{0}]
               e^{-i2\pi (k_{z}z+k_{y}y)}dydz
       \right|^{2}.
\end{equation}
Here, $k_{y}$ and $k_{z}$ are wavenumbers 
(defined as $k=1/\lambda$, where $\lambda$ 
is the wavelength) in the $y$- and 
$z$-directions, respectively. The intensity 
increases and fades when approaching and 
departing a resonance layer, respectively. 
The changes in $f_{i}$ and $|{\bf J}|/J_{0}$ 
along the $x$-axis are demonstrated in five 
slices at 
$x = -0.5\delta, \> -0.3\delta, \> 0, \> 
      0.3\delta, \>  0.5\delta$ 
(see middle and lower panels of 
Fig.$~$\ref{fig05}). The black dash-dotted 
lines in the middle panels indicate the 
vectors perpendicular to the local magnetic 
field. On certain resonance layer 
($x = \pm 0.3\delta$), the line coincides 
with the maximum signal since it must satisfy 
${\bf k} \cdot {\bf B} = 0$, where 
${\bf k} = k_{y}{\bf e_{y}}+k_{z}{\bf e_{z}}$. 
As $B_y$ changes sign across $x = 0$, the 
dominant ${\bf k}$ in the resonance layers 
reverses the sign of one of its component 
while preserving another. The blue solid 
line in the upper panel of Fig.$~$\ref{fig05} 
plots the maximum of $f_i$ on each $zy$-plane 
along $x$-direction, $\max_{k_z,k_y}(f_i)$, 
as a function of $x$. We find that 
$\max_{k_z,k_y}(f_i)$ peaks several times 
inside the primary current sheet. 
Comparisons with the $q$ curve reveal that 
these peaks correspond to resonance layers 
with $q = 1$ and $q = 2$ (red and orange 
dashed lines respectively in upper panel 
of Fig.$~$\ref{fig05}). Resonance layers 
with other $q$ values are missing in this 
model, possibly because that the large-$q$ 
resonance layers near the center ($x = 0$) 
are separated by less than the spatial 
resolution of the simulation. It is also 
likely that the most unstable tearing mode 
in this configuration is the 3D mode 
($k_{z} \neq 0$) rather than a 2D mode 
($k_{z} = 0$), and the modes on layers 
with $q \geqslant 3$ grow at smaller rates 
than the modes on layers with $q = 1$ and 
$q = 2$ \citep{baa12}. 

\begin{figure}[!h]
    \centering
    \includegraphics[scale=0.6]{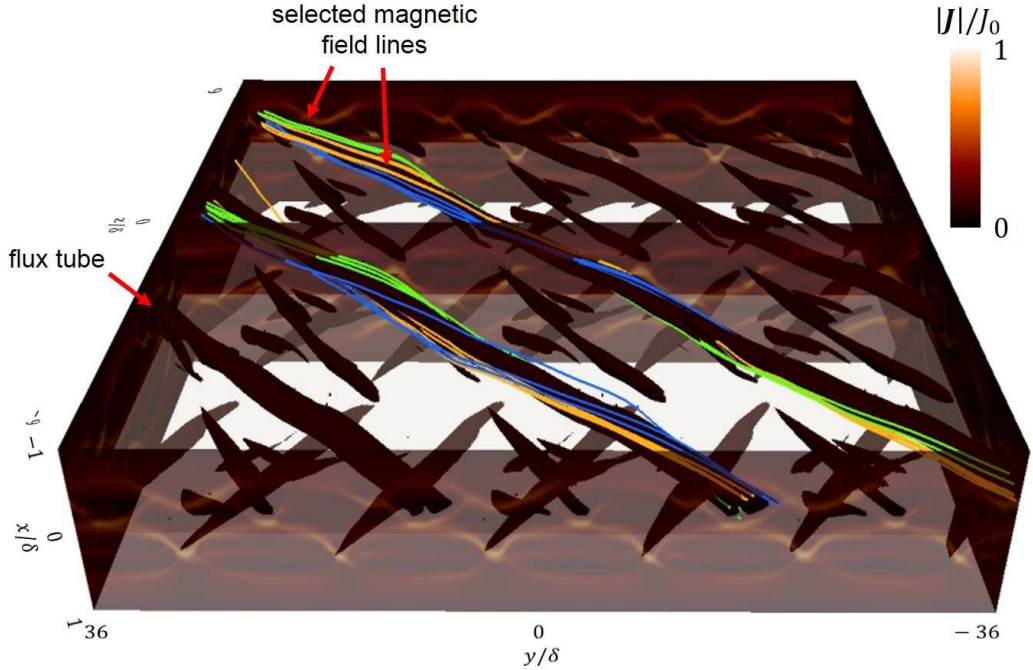}
    \caption{Coherent structure of flux tubes 
             extracted from the isosurfaces of 
             current density 
             $|{\bf J}|/J_{0} = 0.12$ at $t = 500t_{A}$.
             The plotting area in the $y$- and 
             $z$-directions is expanded by a 
             factor of $3$. Translucent white 
             surface represents the $x = 0$ 
             plane. Colored solid lines are 
             selected magnetic field lines.}
    \label{fig06}
\end{figure}

As the guide field is uniform, 
no null points or sheets exist. The magnetic 
field lines reconnect across the tearing layers 
and only part of the field strength is 
consumed. These component reconnections lead 
to twisting reconnected field lines that cross 
the resonance layer to and fro in 3D space. 
The field lines form coherent flux tubes whose 
width along the $zy$-plane approximately equals 
the wavelength $\lambda$ of the tearing mode 
($\lambda = 2.9\delta$ and $5.4\delta$ on 
$q = 1$ and $q = 2$ planes, respectively). 
The flux tube structure can be extracted 
from the isosurfaces of the current density, 
as shown in Fig.$~$\ref{fig06} (since 
periodic boundary conditions are assumed, the 
plotting area in the $y$- and $z$-directions 
is extended by a factor of $3$). Twisting 
field lines wind around the flux tubes. 
The tilting angle of the flux tubes reverses 
across the $x = 0$ plane (indicated by the 
translucent white surface in Fig.$~$\ref{fig06}).

Later, as the local reconnection events 
gradually develop in the individual diffusion 
regions, the flux tubes thicken along the 
$x$-direction. The outflows from the diffusion 
regions are sufficiently accelerated to distort 
the well-organized structure 
(Fig.$~$\ref{fig02}(c)). The flux tubes 
centered at the negative-$x$ side collide and 
coalesce (Fig.$~$\ref{fig02}(d)). Finally, only 
spatially extended current sheets and large flux 
tubes remain (Fig.$~$\ref{fig02}(e)).

\begin{figure}[!h]
    \centering
    \includegraphics[scale=0.85]{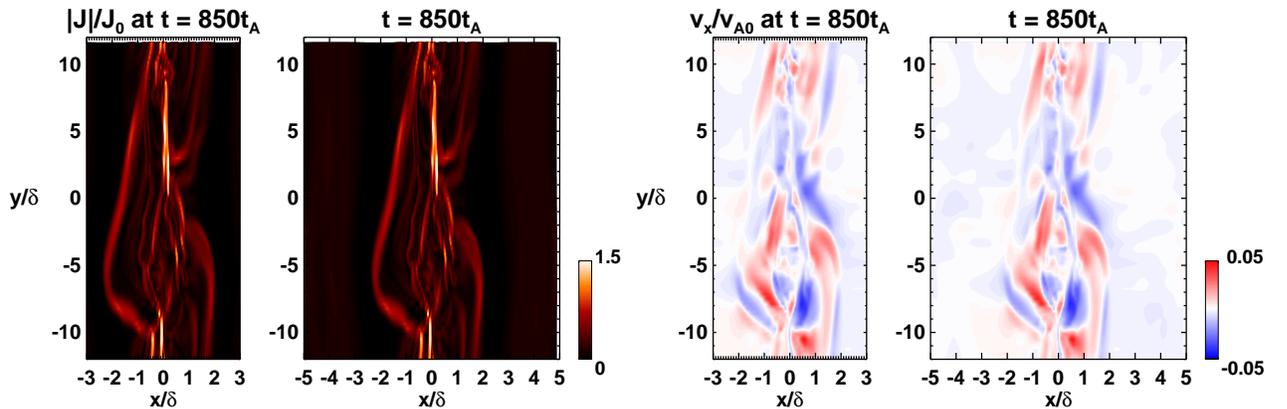}
    \caption{Current density $|{\bf J}|/J_{0}$ 
             and velocity $v_x/v_{A0}$ 
             in the $z = 0$ plane by comparing 
             plot ranges from 
             $-3\delta$ to $3\delta$ 
             and plot ranges from 
             $-5\delta$ to $5\delta$.}
    \label{fig07}
\end{figure}

Actually there are three current sheets 
coexisting inside the whole simulation box. 
It could be seen clearly that the current 
sheets at the boundary have no significant 
influence on the current sheet at the center 
(Fig.$~$\ref{fig07}). Furthermore, We test 
a reconnection model with twice the length 
along $x$-direction to see the effect of 
the boundary and the result does not change.

\subsection{Reconnection rate}

\begin{figure}[!h]
    \centering
    \includegraphics[scale=1.0]{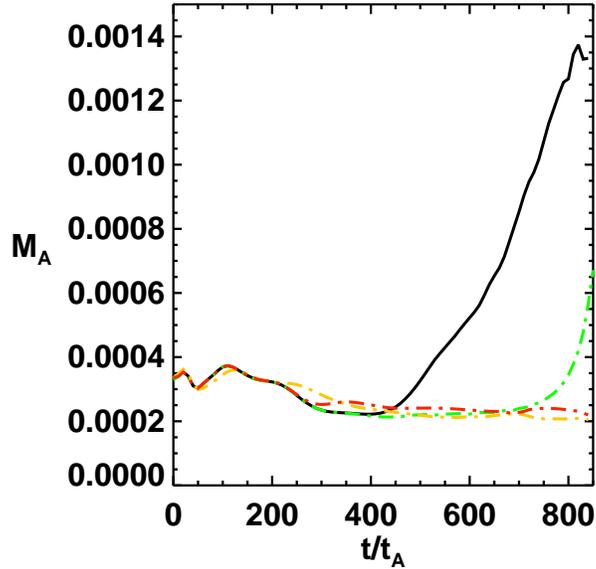}
    \caption{Reconnection rate in the basic 
             model (black solid line), 
             whose maximum is about 5 times 
             of the Sweet-Parker type 
             reconnection simulated in the 
             same 3D box (not plotted).
             2.5D simulation with the guide 
             field (green dash-dotted line), 3D 
             simulation without a guide field 
             (orange dash-dotted line), and 3D 
             simulation with a single
             mode (red dash-dotted line).}
    \label{fig08}
\end{figure}

To understand the reconnection efficiency, 
we acquire the reconnection rate calculated 
as follows:
\begin{equation}
M_{A}=\left| \frac{d}{dt} \int \epsilon_{m} dV 
      \right| {\Big /} \left( 2L_{y}L_{z} 
         \frac{B_{y0}^{2}}{4\pi} v_{A0}\right).
\end{equation}
In the numerator of Eq.$~$(12), the magnetic 
energy density $\epsilon_{m}$ is integrated 
over the volume inside the primary global 
current sheet to obtain the total magnetic 
energy. The boundary of the primary current 
sheet is determined from the specified 
critical plasma pressure 
$p_{c} \sim 0.008 B_{0}^{2}$. 
The energy inside the considered sheet is 
conserved by adding the surface flux term 
along the $x$-direction in Eq.$~$(4). 
The result is plotted in Fig.$~$\ref{fig08}.
The reconnection rate is rapidly enhanced 
after $t = 450t_{A}$.

To check whether 3D reconnection is 
sufficiently fast, we simulated the 
Sweet-Parker type reconnection in the same 
3D box. This model begins with four large 
vortices imitating the inflow-outflow 
coupling in a single reconnection. The 
velocity components are constrained by 
$v_{x}, v_{y} \lesssim 2 \times 10^{-4}v_{A0}$
and $v_{z} = 0$. Translational invariance 
is maintained at the beginning. The Sweet-Parker 
reconnection rate, which is not plotted in 
Fig.$~$\ref{fig08}, is $0.00026$, which is 
approximately $1/5$ the reconnection rate 
of the basic model.

To better quantify the effect of the third 
dimension on the reconnection, we conduct 
a 2.5D simulation with a guide field 
(green dash-dotted line in Fig.$~$\ref{fig08}). 
The initial velocity field is the initial 
velocity field on the $z = 0$ plane in the 
basic model. Because $k_{z} = 0$ 
(or $L_{z} \rightarrow \infty$) throughout 
the simulation box, the resonance layer is 
limited in $x = 0$ plane. A secondary 
instability (plasmoid instability) starts to 
develop at the end of this simulation, and 
the reconnection rate increases later than 
in the basic model. In comparison, no finer 
filamentary structures are detected in the 
basic model. Then the reconnection rate 
increase in the basic model is due to a 
secondary instability other than plasmoid 
instability.

To generate oblique tearing layers, we also 
require a finite guide field. If the initial 
magnetic field has no $z$-component 
($\alpha = 0$), the tearing mode can grow 
only along the $x = 0$ plane, across which 
the magnetic field reverses. The result of 
a 3D simulation without the guide field 
is plotted as the orange dash-dotted line 
in Fig.$~$\ref{fig08}. In this model, the 
initial velocity is randomly perturbed as 
in the basic model. Clearly, the reconnection 
is much lower in this simulation than in 
the basic model.

As shown in the previous result, the basic 
model admits several tearing layers 
coexisting in the same current sheet. To 
understand the importance of these multiple 
layers, we conduct a 3D simulation with 
a single mode (${\bf k}$) in the perturbed 
velocity field (red dash-dotted line in 
Fig.$~$\ref{fig08}). The selected ${\bf k}$ 
is the most unstable mode in the leftmost 
resonance layer (at the negative $x$-side) 
of the basic model. The reconnection rate 
in this model is $\sim 20\%$ that of the 
basic model throughout the same development 
period. The reconnection within each 
diffusion region gradually saturates rather 
than largely increases. This result confirms 
that the reconnection rate is enhanced by 
multiple resonance layers coexisting in one 
current sheet.

\begin{figure}[!t]
    \centering
    \includegraphics[scale=1.0]{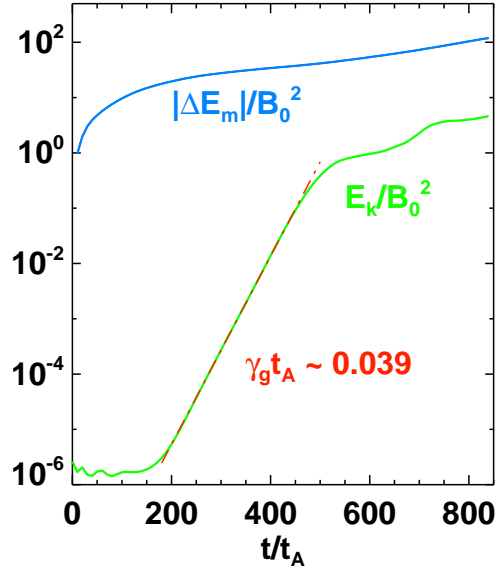}
    \caption{Development of total consumed 
             magnetic energy (blue solid line) 
             and total kinetic energy (green 
             solid line). Red dash-dotted line 
             represents exponential growth.} 
    \label{fig09}
\end{figure}

To quantitatively examine the energy 
conversion efficiency, we plot the temporal 
evolutions of the normalized total kinetic 
energy $E_{k}/B_{0}^2$, and the consumed 
total magnetic energy 
$|\Delta E_{m}|/B_{0}^2$ of the central 
global current sheet in Fig.$~$\ref{fig09}. 
The kinetic energy grows exponentially 
from $t = 200t_{A}$ to $t = 450t_{A}$. From 
exponential fitting, we find that 
$\gamma_{g} t_{A} \sim 0.039$, 
where $\gamma_{g} = d(\ln E_{k})/dt$. 
Comparing these results with the reconnection 
rate, it is easily seen that rapid enhancement 
immediately follows the exponential growth of 
kinetic energy (at $t = 450t_{A}$). Moreover, 
the tearing layers are already recognizable 
around this time (see Fig.$~$\ref{fig02}(b)). 
Together with the above findings, this implies 
that the interaction between fully grown 
tearing layers triggers the fast consumption 
of magnetic energy.

Regarding the different reconnection rates 
plotted in Fig.$~$\ref{fig08}, we conclude 
that the interaction between different 
resonance layers is crucial for the enhanced 
reconnection in later phase.

\subsection{Positive-feedback system}

\begin{figure}[!h]
    \centering
    \includegraphics[scale=1.2]{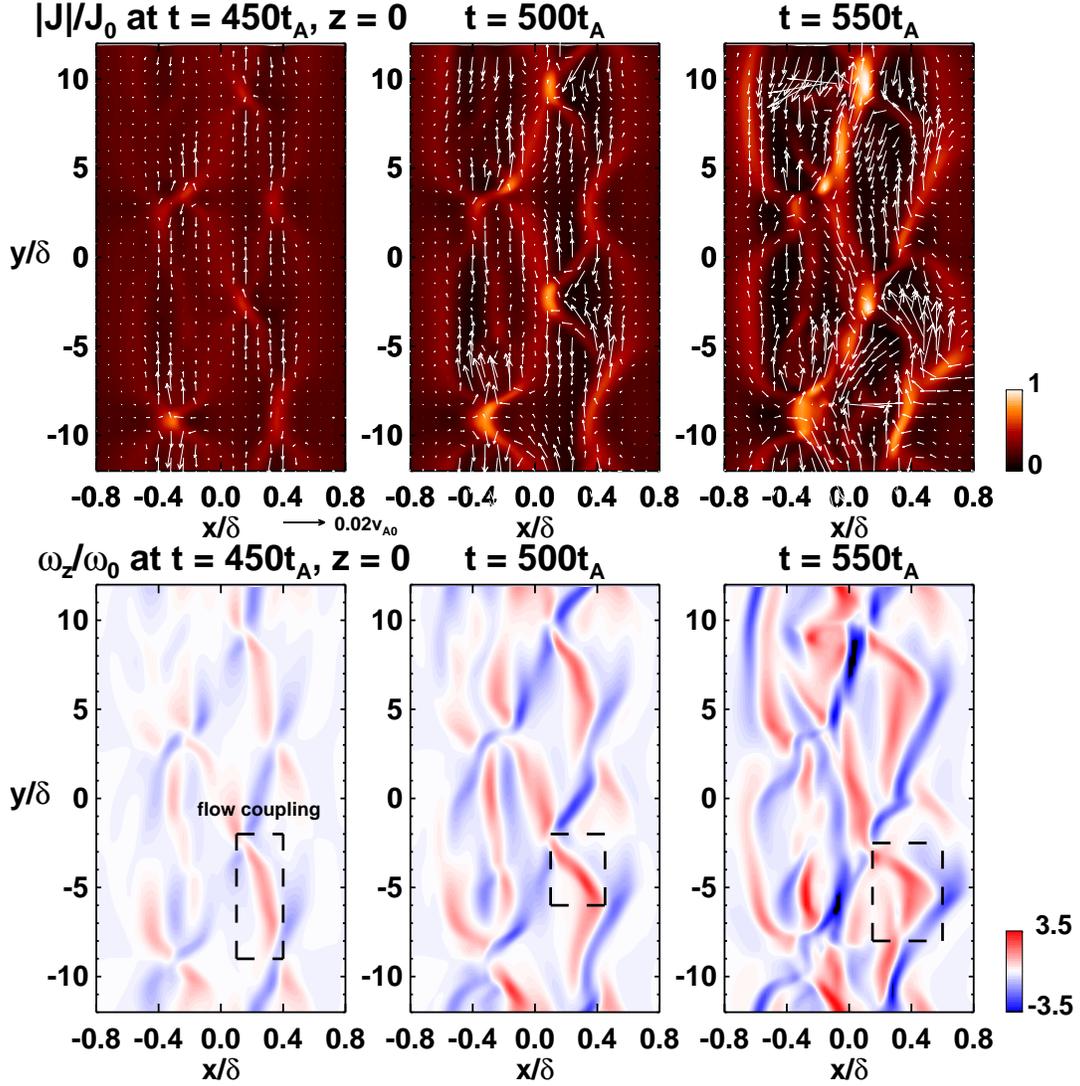}
    \caption{Snapshots of current density 
             $|{\bf J}|/J_{0}$ and the 
             $z$-component of vorticity 
             $\omega_{z}/\omega_{0}$ 
             on $z = 0$ plane at $t = 450t_{A}$, 
             $500t_{A}$, and $550t_{A}$.
             White arrows represent the flow 
             patterns. The vector scale 
             is shown at the bottom of the 
             left upper panel.}
    \label{fig10}
\end{figure}

On closer examination, the diffusion regions 
on multiple tearing layers are observed to 
form a web-like pattern across the current 
sheet. Fig.$~$\ref{fig10} plots the current 
density $|{\bf J}|/J_{0}$ on the $z = 0$ 
plane at various times of the simulation 
($t = 450t_A$, $t = 500t_A$, and $t = 550t_A$), 
overlaid with the velocity flows (white arrows) 
and the $z$-component of the vorticity 
$\omega_{z}/\omega_{0}$, where 
$\omega_{0} = v_{A0}/\delta$. On this plane, 
the small diffusion regions form an asymmetric 
structure with a zigzag pattern. Within this 
structure, the diffusion regions on different 
resonance layers are apart with each other 
in $y$-direction. Examining the local stream, 
we find that the outflow from one reconnection 
site diverts and feeds into the inflow region 
of the reconnection site on a different 
resonance layer. Meanwhile, a portion of the 
reconnected magnetic flux is transported, 
where it can again participate in reconnection. 
The transportation is regulated by the outflow 
from the diffusion region. The outflow 
strengthens as the local reconnection proceeds, 
implying that faster flux transportation 
correspondingly enhances the inflow, thus 
accelerating the local reconnection. This 
coupling, named as positive-feedback system, 
is generated by the existence of multiple 
tearing layers, in contrast to the self-feeding 
system inside a current sheet with a single 
reconnection layer reported by \citet{lap08}. 
The coupling of inflow and outflow regions 
across the current sheet can be identified by 
the coalescence of branch-like structures, which 
extend from the individual diffusion regions 
in the contour plots of $\omega_{z}$. An 
example of such coupling is delineated by 
the black dashed rectangle in the lower panels 
of Fig.$~$\ref{fig10}. Although this feature 
is gradually deformed by the turbulence 
developing inside the current sheet, the 
secondary-transportation of magnetic flux is 
maintained, ensuring that reconnection can 
proceed.

\begin{figure}[!h]
    \centering
    \includegraphics[scale=0.75]{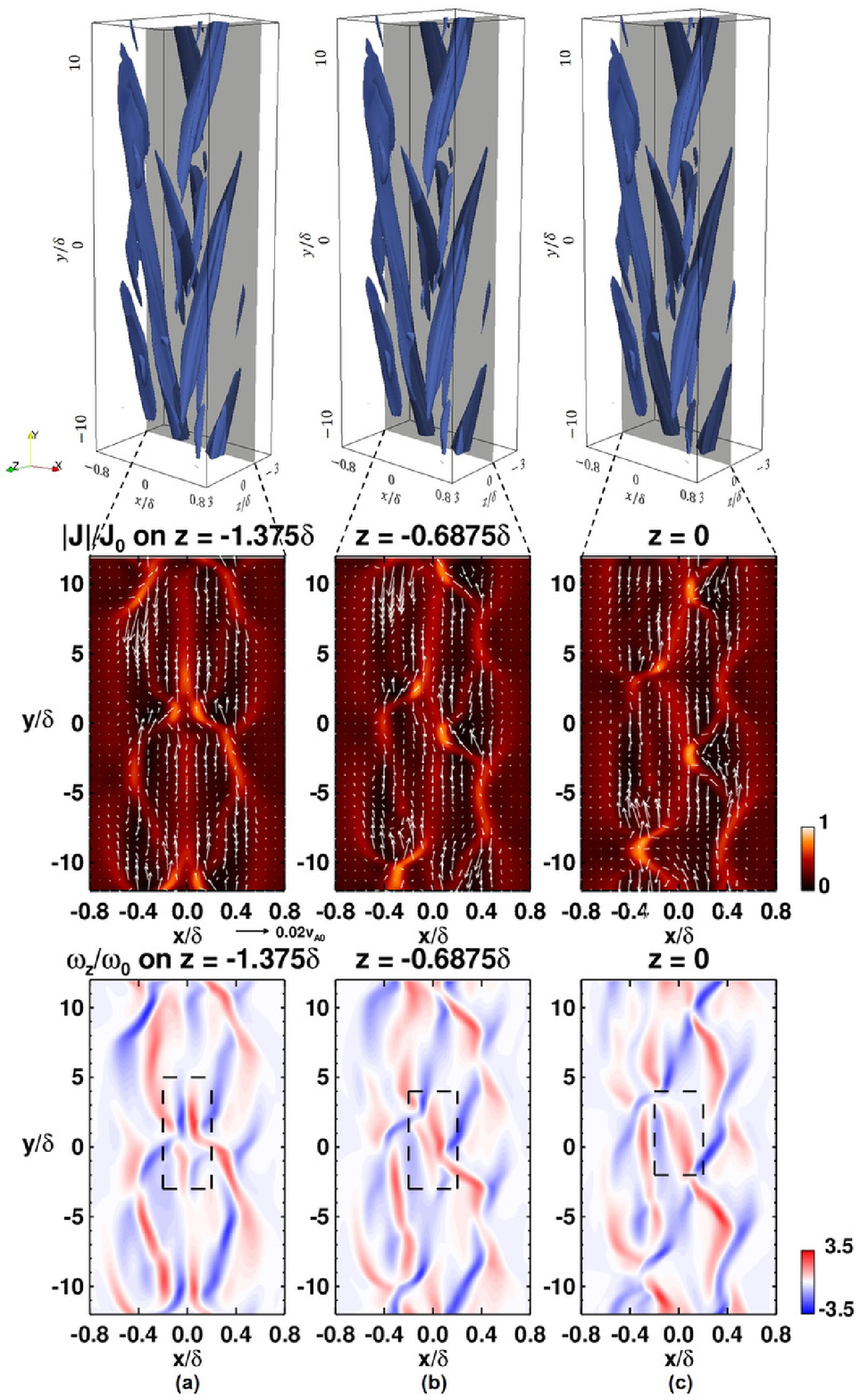}
    \caption{Upper panel: 3D surface of current 
             density $|{\bf J}|/J_{0} = 0.12$ 
             in the simulation box. Black 
             translucent surfaces are the planes 
             at $z = -1.375\delta$, 
             $-0.6875\delta$, and $0$ 
             at $t = 500t_{A}$. 
             Middle panel: Snapshots of 
             $|{\bf J}|/J_{0}$ on the planes 
             shown in the upper panel. White 
             arrows represent the flow patterns.
             Lower panel: Corresponding 
             $\omega_{z}/\omega_{0}$ contour 
             plots.}
    \label{fig11}
\end{figure}

\begin{figure}[!h]
    \centering
    \includegraphics[scale=0.5]{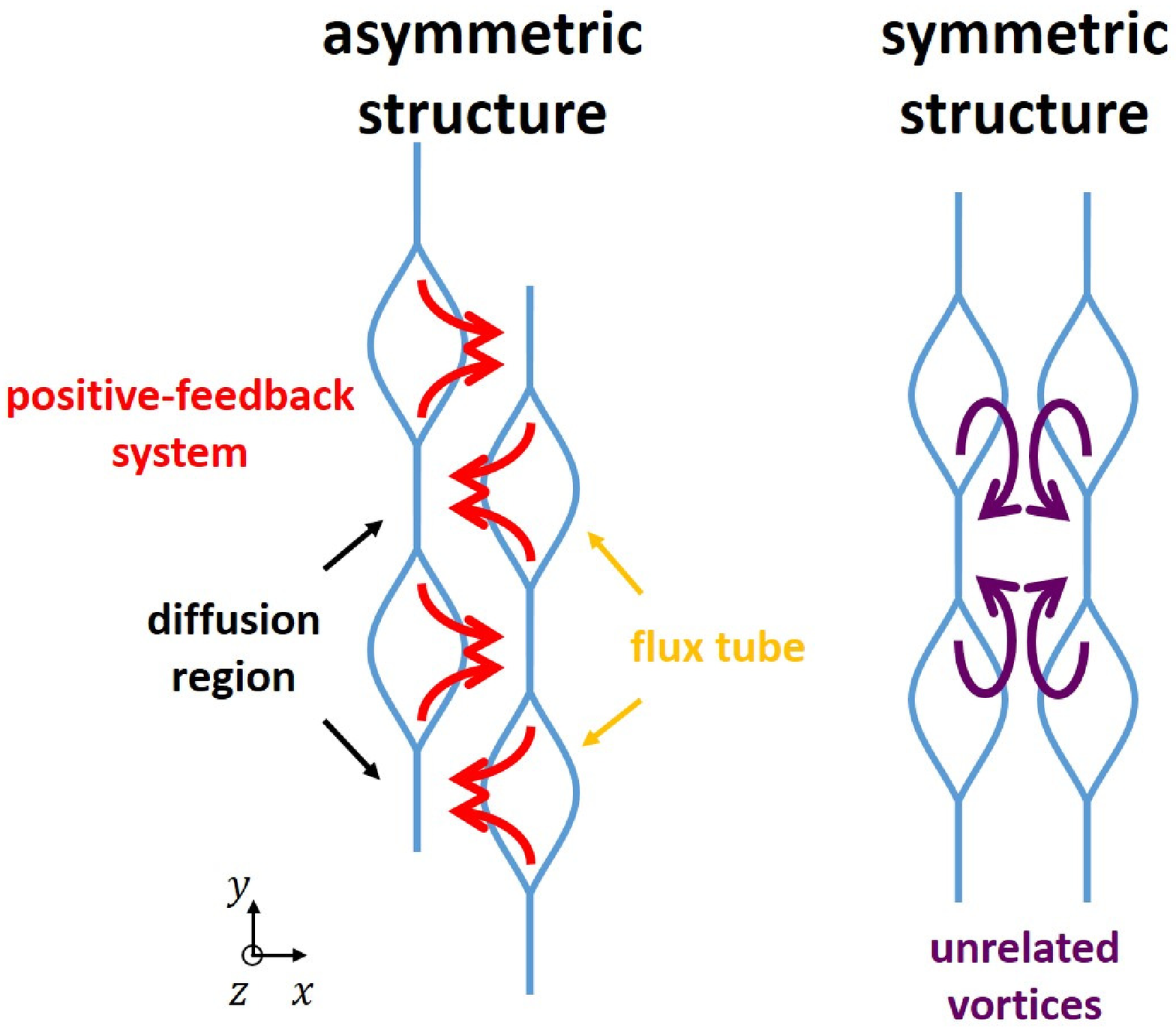}
    \caption{Cartoon of the layout of diffusion 
             regions across the global current 
             sheet. Red and purple arrows 
             represent the characteristic flows 
             on the $xy$-plane.}
    \label{fig12}
\end{figure}

As shown in Fig.$~$\ref{fig05}, if $k_z$ of the 
most unstable modes retain their sign, $k_y$ 
of these modes change sign at opposite side 
of the current sheet center. This implies that 
the tilting angles of the diffusion lines 
(defined as the angles between the orientations 
of the diffusion lines and the $xz$-plane) 
differ at either side of the global sheet, 
as shown in the flux tubes of Fig.$~$\ref{fig06}. 
The tilting angle is defined as:
\begin{equation}
\theta = \arctan{\left( 
            \frac{\langle B_{y} \rangle_x}
                 {\langle B_{z} \rangle_x} 
                 \right)}.
\end{equation}
Therefore, the configuration of the 
diffusion regions changes in different 
$xy$-planes along $z$-axis.  
Fig.$~$\ref{fig11} plots the same variables 
as Fig.$~$\ref{fig10} but at different 
$z$-positions ($t = 500t_A$). Unlike their 
appearance in the $z = 0$ plane, the 
diffusion regions on the different tearing 
layers shift in the $y$-direction 
and sometimes show a symmetric structure 
(in which diffusion regions on different 
resonance layers align along the 
$x$-direction). In this configuration, 
the two diffusion regions do not 
cooperatively interact; thus no feedback 
character is established. The transition 
can be tracked by observing the pattern 
delineated by the dashed rectangle in the 
lower panel of Fig.$~$\ref{fig11}. To an 
observer moving along the $z$-direction, 
the diffusion regions form an alternating 
asymmetric-symmetric-asymmetric structure. 
A cartoon of these structures are 
presented in Fig.$~$\ref{fig12}.

\begin{figure}[!h]
    \centering
    \includegraphics[scale=0.4]{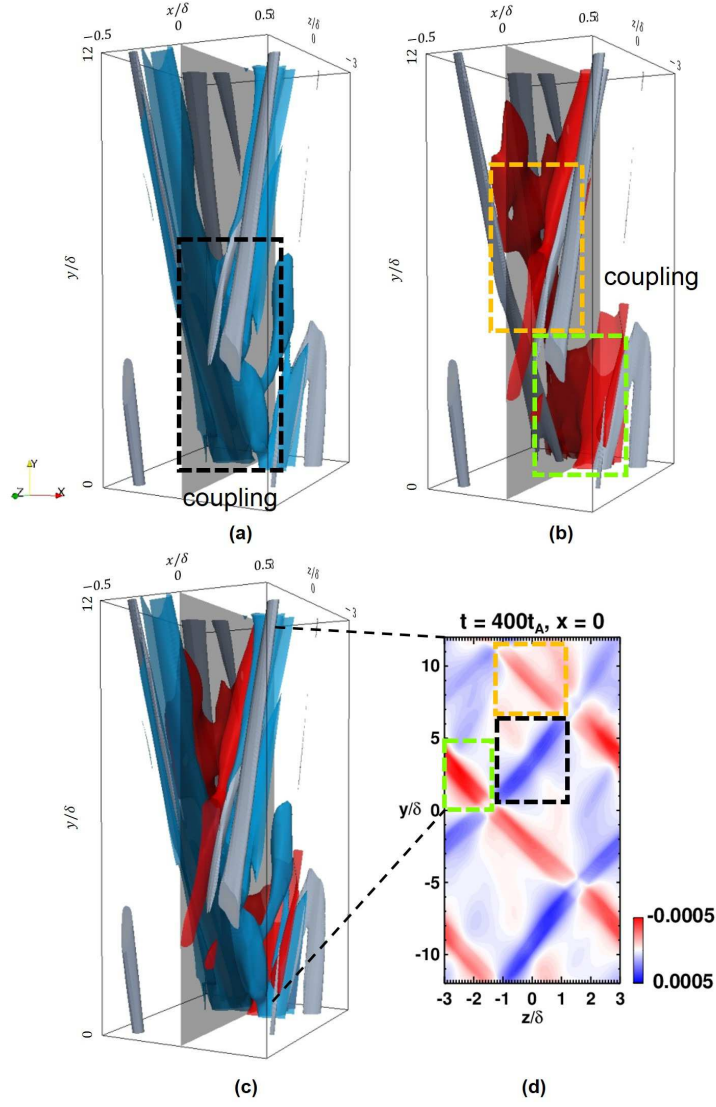}
    \caption{(a)-(c) Surface plots of $v_{x}$ 
             and $|{\bf J}|$ in the 3D box. 
             Grey surface represents 
             $|{\bf J}|/J_{0} = 0.12$. 
             Blue and red surfaces denote 
             $v_{x}/v_{A0} = -0.0005$ and $0.0005$ 
             respectively. Black, green and 
             yellow dashed rectangles highlight 
             the coupling of inflow and outflow 
             regions. (c) combines panels (a) and 
             (b). (d) is the 2D contour plot 
             of $v_{x}$ on the translucent 
             plane in (a)-(c). The same 
             features are highlighted in the 
             the same regions.}
    \label{fig13}
\end{figure}

To examine the global structure of the 
positive-feedback system built by multiple 
tearing layers, we plot the 3D and 2D spatial 
distributions of the local $v_{x}$ 
(see Fig.$~$\ref{fig13}). Once the outflow 
region of a certain reconnection site couples 
with the inflow region of another site, the 
surface of relatively high $v_{x}$ continuously 
extends across the current sheet center. Here, 
we select a cutoff surface of 
$v_{x}/v_{A0} = \pm 0.0005$, approximately 
$20\%$ of the maximum absolute value of 
$v_{x}/v_{A0}$ throughout the box. The dashed 
rectangles in panels (a) and (b) of 
Fig.$~$\ref{fig13} highlight where the surface 
crosses the black translucent plane (the 
central plane of the current sheet at $x = 0$). 
On this plane, the $v_x$ contour plot exhibits 
an oblique checkered pattern, as it is presented 
in Fig.$~$\ref{fig13}(d). The same features are
highlighted in the same regions. Noted that 
$v_{x}$ is not enhanced at the corners of the 
checkered pattern, because the diffusion 
regions are symmetrized at those points. Where 
the diffusion regions become asymmetric, 
the positive-feedback system establishes 
provided that a little shift occurs between the 
diffusion regions. Thus, the $v_x$ is 
distinctly enhanced along the borders of the 
checks.

In summary, the diffusion regions in different 
resonance layers couple via their inflow 
and outflow regions. This configuration, 
the positive-feedback system, globally occurs 
inside the box, and underlies the rapid 
reconnection observed in our model.

\subsection{Global inflow and slow-mode shocks}

\begin{figure}[!h]
    \centering
    \includegraphics[scale=1]{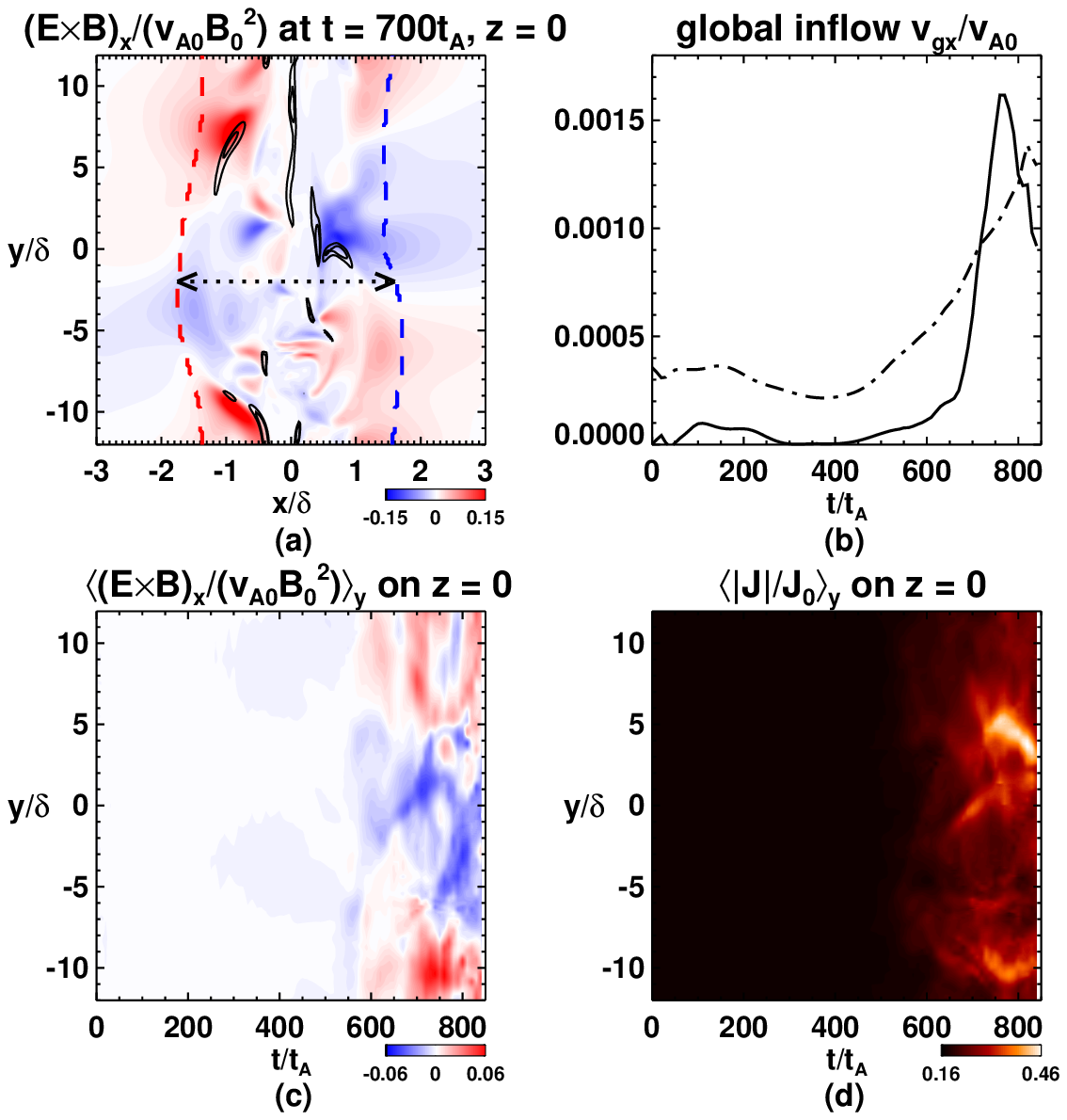}
    \caption{(a) Normalized Poynting flux at 
             $700t_{A}$ on the $z = 0$ plane. 
             Black contours indicate the 
             enhanced current density 
             $|{\bf J}|/J_{0} \ge 0.8$.
             Current sheet boundary 
             ($p_{c} = 0.008B_{0}^{2}$) is 
             marked by the red and blue dashed 
             lines to the left and right of 
             $x = 0$, respectively. 
             (b) Averaged global inflow (black 
             solid line) and reconnection 
             rate (black dash-dotted line). 
             (c) and (d) Spatially averaged 
             normalized $({\bf E \times B})_{x}$ 
             and $|{\bf J}|$ inside the current 
             sheet indicated by the dotted 
             double arrows in (a).}
    \label{fig14}
\end{figure}

The enhanced reconnection rate in the basic 
model implies that a large amount of 
magnetic energy is transported into the 
current sheet. To evaluate the rate of 
energy transport, we investigate the global 
inflow along the $x$-boundary of the central 
current sheet (Fig.$~$\ref{fig14}). 
Fig.$~$\ref{fig14}(a) exhibits snapshot 
of the normalized Poynting  pattern 
${(\bf E \times B)}_{x}$ at the 
time of Fig.$~$\ref{fig02}(d) (namely, 
at $t = 700t_{A}$). The black contours 
indicate where the current density is locally 
enhanced. The surfaces assumed in the average 
inflow calculation are highlighted by the 
colored dashed lines to the left and right 
of the global current sheet. They are the 
same surfaces used in the reconnection rate 
calculation. The average global inflow is 
determined as follows: 
\begin{equation}
v_{gx} = \frac{\oint ({\bf E} \times {\bf B}) \cdot {\bf dS}}
         {2{B_{0}^{2}L_{y}L_{z}}}
\end{equation}
where ${\bf E}$ is the electric field and 
${\bf dS}$ denotes the surface on global 
current sheet boundary. We consider the dot 
product to be only the $x$-component of 
${\bf E} \times {\bf B}$, thus 
$|{\bf dS}| = dydz$. This quantity reflects 
the transport rate of the Poynting flux into 
the current sheet in the $x$-direction. As 
shown in Fig.$~$\ref{fig14}(b), the $v_{gx}$ 
changes smoothly over time. Panels (c) and (d) 
of Fig.$~$\ref{fig14} present time profiles 
of the normalized ${\bf E \times B}$ and 
$|{\bf J}|$, respectively. Both variables 
are spatially averaged inside the current 
sheet as follows (where $f$ denotes a 
variable)
\begin{equation}
\langle f \rangle_{y} = 
 \frac {{\int_{x_{l}(y)}^{x_{r}(y)} f(x,y,z=0) dx}}
       {x_{r}(y)-x_{l}(y)}.
\end{equation}
In Eq.$~$(15), $x_{l}(y)$ and $x_{r}(y)$ 
denote the positions of the left and right 
boundaries, respectively, of the current 
sheet at $(y, z = y, 0)$. The diffusion 
region is asymmetric in the $z = 0$ plane 
at $t = 700t_A$, as observed in 
Fig.$~$\ref{fig14}(a). Consequently, the 
inflow regions extending from the current 
sheet are also asymmetric. This pattern 
becomes increasingly obvious as local 
reconnection is accelerated by 
positive-feedback system (Fig.$~$\ref{fig14}(c)). 
When entering the turbulent state (after 
$\sim t = 600t_{A}$; see Fig.$~$\ref{fig14}(d)), 
the dense regions essentially overlap with 
the structures of Fig.$~$\ref{fig14}(a) and (c), 
and the current density inside the current 
sheet is universally enhanced. Although 
the inflow structure is segmented, the global 
effect can be regarded as a single long 
diffusion region. Moreover, beyond 
$t = 400t_A$, the $v_{gx}$ well correlates 
with the reconnection rate (black dash-dotted 
line in Fig.$~$\ref{fig14}(b)). The enhanced 
$v_{gx}$ suggests that, on macroscopic scale, 
magnetic energy is carried into the global 
current sheet by the flow and converted into 
other forms of energy. This suggestion is 
supported by the rising reconnection rate in 
later phase. The average global inflow 
exceeds the reconnection rate at approximately 
$t = 700t_{A}$, indicating that a small 
portion of the magnetic energy is stored 
before being gradually consumed. 

\begin{figure}[!h]
    \centering
    \includegraphics[scale=1.2]{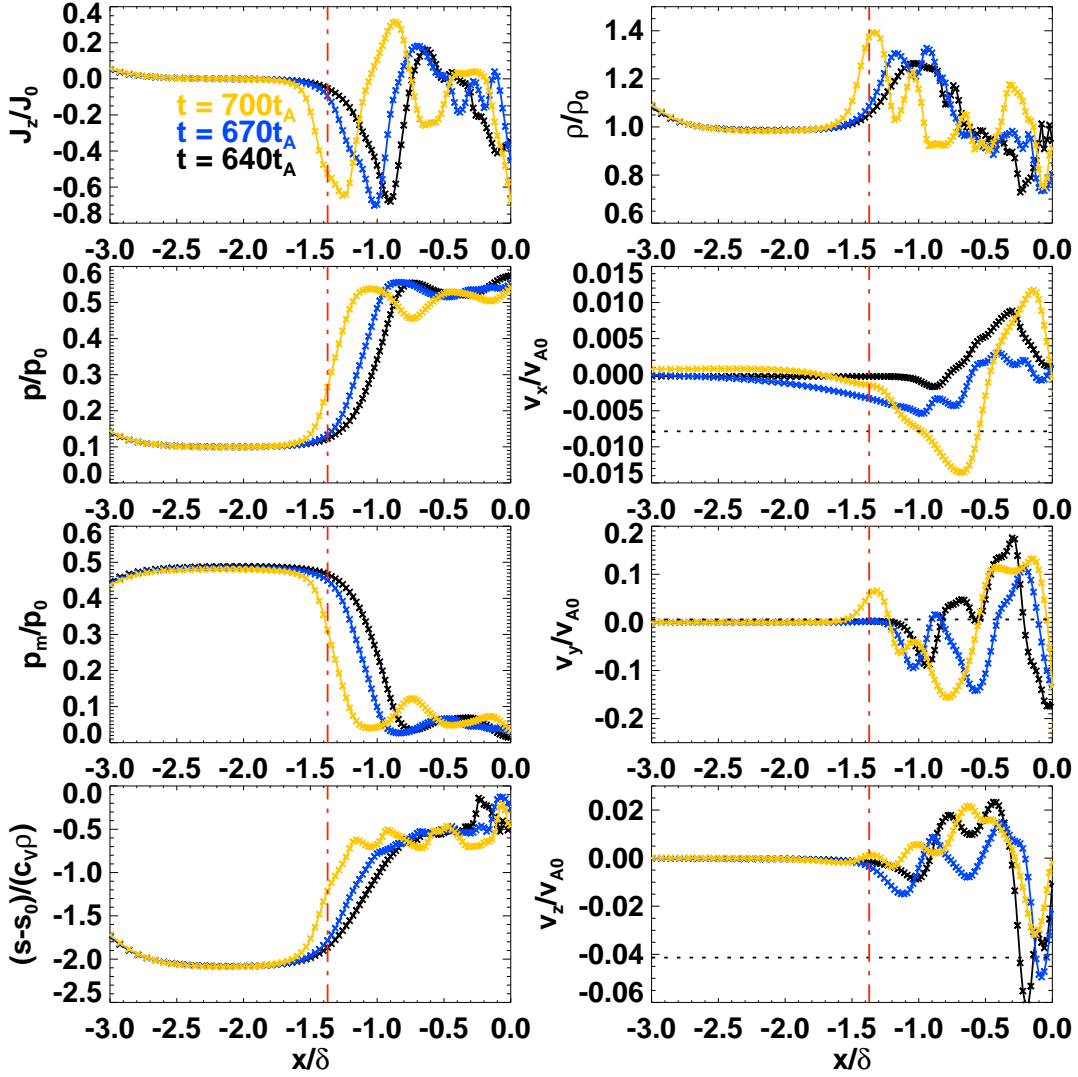}
    \caption{Plots of normalized current density, 
             plasma pressure, magnetic pressure, 
             entropy, plasma density and velocity 
             across $(y,z = 0,0)$, showing 
             the development of a slow-mode 
             shock. Variables are plotted at 
             $t = 640t_{A}$ (black solid lines), 
             $670t_{A}$ (blue solid lines), 
             $700t_{A}$ (orange solid lines).
             Crosses indicate the grid points. 
             Red dash-dotted line is the detected 
             shock front at $t = 700t_{A}$. 
             Black horizontal dotted lines denote 
             the shock speed in lab-frame.}
    \label{fig15}
\end{figure}

When the inflow enhancement is high, slow-mode 
shocks gradually form between the inward flow 
and the outwardly growing flux tube. Along 
with reconnection in the diffusion region, these 
shocks are a likely mechanism of magnetic energy 
conversion. Shocks are identified by comparing 
the upstream and downstream quantities in the 
shock frames at selected points. Slow-mode 
shocks in the simulation domain are required to 
satisfy the following criteria: \\
1. Rankine-Hugoniot relations: The 
Rankine-Hugoniot relations in the downstream 
must not deviate by more than 30\% of their 
upstream values \citep{sai95}; \\
2. Velocities must satisfy 
$v_{Au} \geqslant v_{nu} \geqslant v_{slu}$, 
where $v_{nu}$ is the normal component 
(perpendicular to the shock front) of the 
upstream velocity. $v_{Au}$ and $v_{slu}$ 
are the Alfv\'{e}n speed and the slow-mode 
wave phase speed, respectively, in the 
upstream; \\
3. $v_{nd} \leqslant v_{sld}$, where $v_{nd}$ 
denotes the normal component of the downstream 
flow speed and $v_{sld}$ is the slow-mode 
wave phase speed in the downstream; \\
4. The ${\bf B}$ and ${\bf v}$ of the upstream 
and downstream must be co-planar 
(within $10\degr$). 

Fig.$~$\ref{fig15} shows the development of 
a slow-mode shock wave. Plotted are the 
current density $J_{z}/J_{0}$, plasma pressure 
$p/p_{0}$, magnetic pressure $p_{m}/p_{0}$, 
entropy $(s-s_{0})/c_{V}$, 
plasma density $\rho/\rho_{0}$, 
plasma velocity $v_{x}/v_{A0}$, $v_{y}/v_{A0}$ 
and $v_{z}/v_{A0}$ across $(y,z = 0,0)$ in 
laboratory frame at three time points (for 
tracking the transition). The entropy is 
normalized by 
$s_{0} = c_{V}\ln(p_{0}/\rho_{0}^\gamma)$, 
where $c_{V}$ is the specific heat at constant 
volume. The shock speed in the laboratory frame 
is indicated by the black dotted line in the 
plasma velocity plots. At $t = 640t_A$ (black 
solid lines in Fig.$~$\ref{fig15}), the density 
is compressed around $x = -\delta$. The local 
$J_{z}$ is minimized at the $x$-axial boundary 
of the flux tube. At this moment, the local 
$x$-directional flow on either side of the 
flux tube boundary is relatively weak. As the 
reconnection becomes more efficient, the 
flux tube thickens and expands in the 
negative $x$-direction. The relative flow 
between inside and outside of the tube boundary 
increases, further compressing the local plasma 
(compression occurs around $x = -1.3\delta$). 
Eventually (at $t = 700t_{A}$), the shock 
criteria are reached. The approximate position 
of the shock front is indicated  as red 
dash-dotted line in Fig.$~$\ref{fig15}. 
We fail to find the rotational discontinuity 
previously reported in reconnection with a 
guide field \citep[e.g.,][]{lon09}.

\begin{figure}[!h]
    \centering
    \includegraphics[scale=1.0]{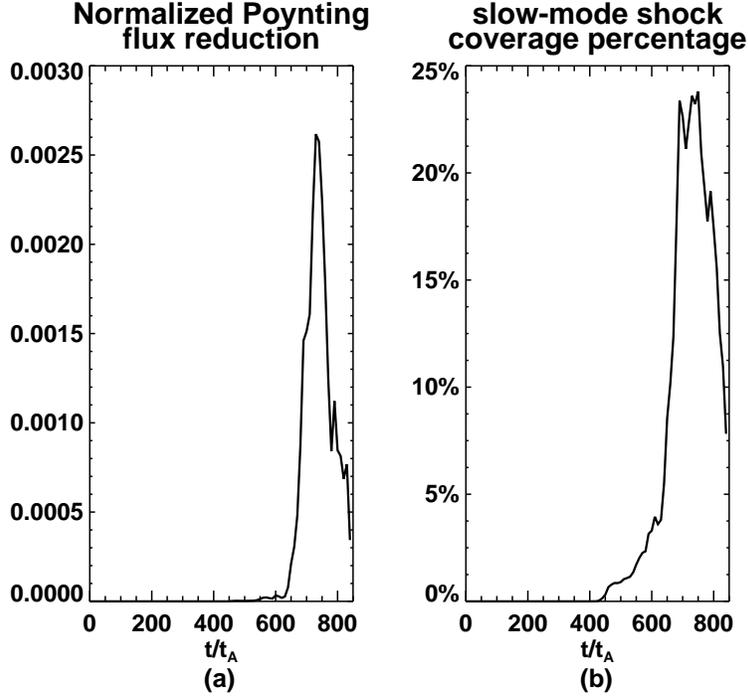}
    \caption{Temporal evolutions of 
             (a) normalized energy conversion 
             capability of shock $M_{s}$ 
             between the upstream and 
             downstream of slow-mode shocks 
             throughout the simulation box, 
             and (b) percentage area coverage 
             of slow-mode shock.}
    \label{fig16}
\end{figure}

\begin{figure}[!h]
    \centering
    \includegraphics[scale=0.45]{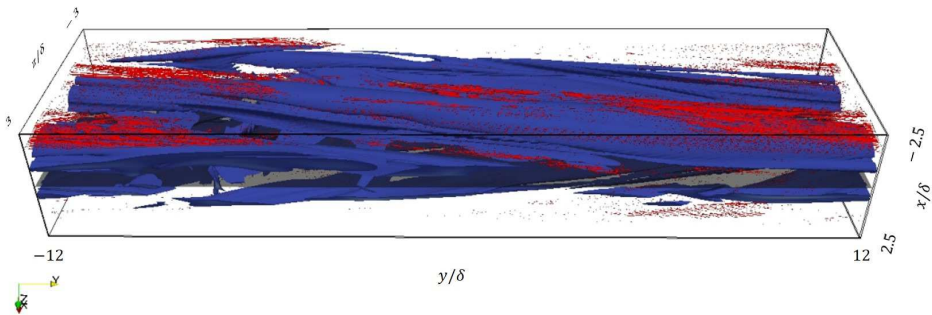}
    \caption{All detected shocks (indicated by 
             red dots) are located approximately 
             by downstream position at $t = 700t_{A}$. 
             Blue surface represents $J_{z} = 0$.}
    \label{fig17}
\end{figure}

Along with local Ohmic heating, these slow-mode 
shocks are considered as an additional source 
of magnetic energy consumption, as shown in 
Petschek's work. The energy conversion 
capability of shock (reduction of Poynting 
flux across the shock) $M_{s}$ is calculated as:
\begin{equation}
M_{s} = \frac{\sum [({\bf E} \times {\bf B})|_{u}
                   -({\bf E} \times {\bf B})|_{d}]
              \cdot dS}{2B_{0}^{2}L_{y}L_{z}}.
\end{equation}
The first and second terms of numerator denote 
the upstream and downstream Poynting vectors, 
respectively. The $M_{s}$ is plotted as a 
function of time in Fig.$~$\ref{fig16}(a). 
The efficiency of these shocks is greatly 
increased at the end of the simulation. 
To understand the global layout of the slow-mode 
shocks, we calculate the total percentage 
area coverage SC of the shock:
\begin{equation}
\textrm{SC}=\frac{S_{ss}}{4L_{y}L_{z}} \times 100\%
\end{equation}
where $S_{ss}$ is the total area covered by 
the shock. The coefficient ``4'' in the 
denominator indicates that four tearing 
layers coexist in the central current sheet.
The temporal evolution of SC is plotted 
in Fig.$~$\ref{fig16}(b). Both $M_{S}$ and 
SC increase from $t = 450t_A$. Especially 
in the later phase, when the global 
inflow booms (after $t = 650t_{A}$), large 
extent of slow-mode shocks with efficient energy 
conversion capability are detected. The 
approximate downstream positions of all 
shocks found in the domain are highlighted 
in red in Fig.$~$\ref{fig17}. The blue 
isosurface is the flux tube surface 
with $J_{z} = 0$.

In summary, multiple tearing layers construct 
a positive-feedback system that promotes 
reconnection. Large amount of magnetic energy 
is transported into the current sheet and 
converted to other forms of energy. Outwardly 
growing flux tubes collide with the strong 
inflow, gradually arousing slow-mode shocks 
along the tube boundaries. These shocks 
further enhance the reconnection. Therefore, 
we refer to our model as the 
``shock-evoking positive-feedback" model.

\section{Discussion}

In our present model, the reconnection 
rate is enhanced by non-linear interactions 
among multiple (tearing) layers. 
Our observed interactions resemble 
the double tearing mode (DTM) reported in 
the previous studies \citep[e.g.,][]{fur73}. 
Several comparisons are worth a brief 
mention here. First, the unperturbed state 
of DTM has multiple independent current 
sheets, whereas our model begins with 
a single current sheet. The multiple current 
layers are consequent to the growth of 
the tearing layers by mixed perturbations. 
Such growth of multiple layers requires 
a 3D system and a moderate guide field.
Second, DTM is essentially a 2D process 
because translational invariance is assumed 
along the guide field direction 
\citep[e.g.,][]{jan11}. Therefore, the 
arrangement of diffusion regions are 
maintained in that direction. In our model, 
the arrangement of diffusion regions differs 
among $z$-positions (see Section 3.3) and 
periodically changes from symmetric to 
asymmetric. Because of this non-uniformity, 
positive-feedback system is intermittently 
distributed. Our model is expected to 
consume magnetic energy less efficiently 
than DTM. Nevertheless, our model 
substantially enhances the reconnection rate.
Third, as also shown in Section 3.3, the 
energy release may be slow in regions of 
symmetric arrangement of diffusion regions 
from different tearing layers. Notably, 
however, reconnection proceeds in symmetric 
DTM structures albeit very slowly \citep{yan94}. 
This result is consistent with our model.

In applying our model to realistic systems, 
such as solar flares, we must reconsider the 
boundary effects and the resistivity, which 
is much larger in our model than in solar 
phenomena. In our simulation, the selection 
of the tearing layers is highly influenced 
by the box size; in other words, by the 
aspect ratio of the initial current sheet. 
In a much larger system enclosing a current 
sheet with an extremely large aspect ratio, 
the evolution should be influenced by the 
boundary effects. If the plasma $\beta$ is 
high in the current sheet, local Alfv\'{e}n 
speed is slow. The evolution may be locally 
determined and the system will behave as an 
open system with free outflows. In 
low-$\beta$ plasma, the evolution is 
globally determined and influenced by the 
boundary conditions at both ends of the 
field lines. Meanwhile, the most unstable 
tearing mode depends on both the resistivity 
and the guide field. However, we argue that 
once the emerging multiple tearing layers are 
sufficiently close to interact (separated 
within several $1/|\Delta'|$, where $1/|\Delta'|$ 
is the width of tearing layer), a 
positive-feedback system is built. Conversely, 
if the multiple layers approach very closely 
(within $1/|\Delta'|$), they will behave as a 
single layer. These propositions require 
investigation in further study; for instance, 
we should elucidate the scaling law that 
relates the Lundquist number to the 
reconnection rate, survey the guide field 
strength parameter and investigate the 
boundary condition. We must also improve the 
resolution and precision of our simulation.

\begin{figure}[!h]
    \centering
    \includegraphics[scale=0.61]{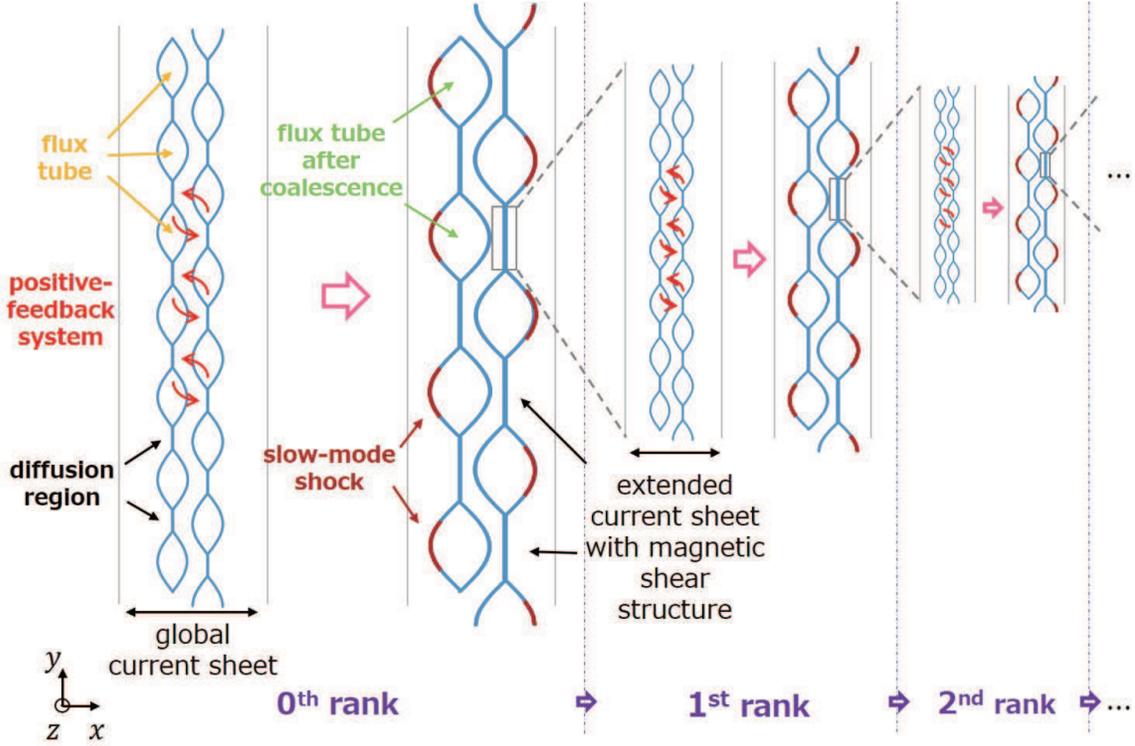}
    \caption{Illustration of hierarchical structure 
             in the ``shock-evoking positive-feedback" 
             model.}
    \label{fig18}
\end{figure}

Finally, we propose an extended model based 
on our ``shock-evoking positive-feedback" 
model, which operates similar to 2D hierarchical 
reconnection. In the later phase of our simulation, 
long and thin current sheets with sheared magnetic 
structures remain near the center, which are 
candidate structures for further tearing. 
\citet{dau11} simulated reconnection in a 
single current sheet under a guide field effect 
in the kinetic regime. They reported secondary 
flux ropes resulting from tearing instability. 
No secondary filamentary structure is observed in 
our model, likely because of the low resolution. 
Higher-ranking filaments are expected in our 
model when the Lundquist number and 
resolution are improved. Once the extended 
current sheet becomes unstable to the tearing 
instability, there are two possibilities. 
If the most unstable mode is centralized, 
a single tearing layer can be identified, 
as occurs in 2D plasmoid instability. 
Conversely, if the most unstable mode is 
oblique \citep{baa12}, multiple tearing layers 
should emerge. The coexisting multiple layers 
resemble the primary current sheet structure. 
Once established, the positive-feedback system 
enhances local reconnection. Flux tubes at the 
same side grow and merge with each other. 
Slow-mode shocks are evoked along the outer 
boundary, while extended current sheets form 
between the coalesced flux tubes. These remaining 
current sheets might be subject to further 
tearing instability. Therefore, in a series 
of steps, the global structural scale of the 
diffusion region could reduce to microscopic 
size (such as the ion inertial scale) by 
alternate applications of 2D plasmoid 
instability and the 3D 
``shock-evoking positive-feedback" model. 
We regard this fractal structure of the 
current sheet generated by the 
``shock-evoking positive-feedback" model as 
an extension of the hierarchical structure of 
2D plasmoid instability. Our scenario is 
illustrated in Fig.$~$\ref{fig18}, assuming 
that multiple tearing layers are realized 
in each rank.

\section{Summary}

In this study, we simulate a 3D MHD magnetic 
reconnection across a symmetric current sheet 
with a finite guide field. We assume a uniform 
resistive environment and randomly perturb 
the initial velocity field. When relaxing the 
variation along the direction of the guide field, 
we find that the guide field increases the 
reconnection rate of the whole current sheet 
by several times, relative to the simulation 
with no guide field. Most unstable tearing 
modes, with components in the sheet-wise 
direction, emerge on multiple resonance layers. 
The diffusion regions on these layers establish 
a zigzag pattern that couple the inflow region 
and outflow region on different layers. 
Consequently, the outflow is diverted into 
the inflow region, ensuring that reconnection 
proceeds in the opposite layer. Gradually, 
the inflow from outside of the global current 
sheet also becomes accelerated by continuous 
activation of individual reconnection site. 
This enhanced inflow arouses slow-mode shocks 
along the outer boundary of the current sheet, 
further promoting energy conversion. We refer 
to our model as the 
``shock-evoking positive-feedback" model.

The Lundquist number is much smaller 
in our numerical experiment than in the solar 
corona. Therefore, our model is not directly 
applicable to real solar activities. 
To understand the feasibility of the 
``shock-evoking positive-feedback" in low 
resistive environments, we must conduct a 
parameter survey on the diffusivity magnitude. 
Altering the resistivity and guide field 
strength would induce different most unstable 
tearing modes, thus changing the topology of 
the positive-feedback system. In future study, 
we will further evaluate the effectiveness of 
our model by varying $B_{z}$ and $\tilde{\eta}$.

The limited resolution of our model precludes 
the detection of finer filamentary structure. 
If the remaining long, thin current sheet 
becomes vulnerable to tearing instability, 
the internal sheared structure might develop 
into a higher-ranking positive-feedback system. 
The mechanism of the 
``shock-evoking positive-feedback" model 
would then enhance the local reconnection rate. 
In this way, the pattern could be repeated on 
ever smaller scales, eventually reaching the 
microscopic scale in environments of large 
Lundquist number. Therefore, we have 
potentially expanded 2D fractal reconnection 
into a 3D hierarchical structure with a high 
energy conversion rate. This capability of 
our model needs to be tested on very fine 
simulation grids, another goal of our future 
work.

\acknowledgments
This research was conducted using the Fujitsu 
PRIMEHPC FX10 System (Oakleaf-FX,Oakbridge) in 
the Information Technology Center, The University 
of Tokyo.
Numerical computations were in part carried out on 
Cray XC30 at Center for Computational Astrophysics, 
National Astronomical Observatory of Japan.
This research is supported by Leading Graduate 
Course for Frontiers of Mathematical Sciences 
and Physics (FMSP) and JSPS KAKENHI Grant Number 
15H03640.

\clearpage

\end{document}